\documentstyle[epsfig,aps]{revtex}
\begin{document}
\draft
\title{Electronic structure of Co$_x$TiSe$_2$ and Cr$_x$TiSe$_2$}
\author{A.~N.~Titov, A.~V.~Kuranov, and V.~G.~Pleschev} 
\address{Ural State University, Physics of Condensed Matter Department,
620083 Yekaterinburg, Russia}
\author{Yu.~M.~Yarmoshenko and M.~V.~Yablonskikh}
\address{Institute of Metal Physics, 
Russian Academy of Sciences-Ural Division, 
620219 Yekaterinburg GSP-170, Russia}
\author{A.~V.~Postnikov}
\address{Theoretical Low-Temperature Physics, 
University of Duisburg, D-47048 Duisburg, Germany}
\author{S.~Plogmann and M.~Neumann}
\address{Department of Physics, Osnabr\"uck University, 
D-49069 Osnabr\"uck, Germany}
\author{A.~V.~Ezhov and E.~Z.~Kurmaev}
\address{Institute of Metal Physics, 
Russian Academy of Sciences-Ural Division, 
620219 Yekaterinburg GSP-170, Russia}
\date{\today}
\maketitle
\begin{abstract}
The results of investigations of intercalated compounds
Cr$_x$TiSe$_2$ and Co$_x$TiSe$_2$ by 
X-ray photoelectron spectroscopy (XPS) and 
X-ray emission spectroscopy (XES) are presented. The data obtained 
are compared with theoretical results of spin-polarized 
band structure calculations. 
A good agreement between theoretical and experimental data 
for the electronic structure of the investigated materials has been observed. 
The interplay between the $M3d$--Ti$3d$ hybridization
($M$=Cr, Co) and the magnetic moment at the $M$ site is discussed.
A 0.9 eV large splitting of the core Cr$2p_{3/2}$ level was observed,
which reveals a strong exchange magnetic interaction of 3d-2p electrons of Cr. 
In the case of a strong localization of the Cr$3d$ electrons (for $x<0.25$), 
the broadening of the Cr$L$ spectra into the region of the states
above the nominal Fermi level was observed and attributed to X-ray re-emission. 
The measured kinetic properties are in good accordance with spectral 
investigations and band calculation results.
\end{abstract}

\section{Introduction}
\label{sec:intro}

Intercalated compounds built by insertion of transition metal (TM) atoms 
into layered titanium dichalcogenides are interesting as natural 
nanostructures formed by alternation of magnetic TM layers 
and non-magnetic layers of the host lattice. The investigations 
performed for $M_x$TiS$_2$ ($M$ = Mn, Fe, Co, Ni) 
\cite{PSSb132-295,PSSb138-357,PRB38-3676} 
show that the effect of intercalation on the chemical bonding 
is not consistent with the rigid band model \cite{AdvPh36-1}
often referred to in relation to these compounds. Instead,
the charge carriers localization due to the formation of covalent
bonds with intercalant atoms seems to become important.
From this point of view, the TiSe$_2$-based intercalated materials 
are obviously even more interesting, because they demonstrate a higher degree 
of two-dimensionality in comparison with TiS$_2$, as follows
from their larger $c/a$ ratio (1.697 in pure TiSe$_2$ vs.
1.672 in TiS$_2$, see, e.g., Ref.~ \onlinecite{Hibma82}).
Tazuke {\em et al.\/} \cite{TaTa_JPSJ97} discussed magnetic properties of 
$M_x$TiSe$_2$ ($M$ = V, Cr, Mn, Fe, Co, Ni, Cu) referring to 
the hybridization between the intercalant $3d$ states and the
band states of TiSe$_2$, the information on the latter
being based on the band structure calculation \cite{YSM_JPC87}.
According to Ref.~\cite{YSM_JPC87}, the hybridized states 
with lighter TM atoms have smaller contribution of Se$4p$ states.

The aim of the present work is to deepen the understanding
of the chemical bonding and related (magnetic and transport)
properties of two TiSe$_2$-based systems, intercalated
with ``light'' (Cr) and ``heavy'' (Co) TM atoms.
Monocrystalline samples over a range of intercalant
concentration were prepared and used in the study of magnetic
susceptibility, conductivity, X-ray emission and 
photoelectron spectroscopy.  

While for the system Co$_x$TiSe$_2$ the literature data about 
influence of intercalation on crystal structure \cite{JSSC17-9}, 
electric and magnetic properties \cite{FTT39-1618} 
were reported earlier, 
for Cr$_x$TiSe$_2$ system we failed to find any literature data except
those of Ref.~\cite{TaTa_JPSJ97}.

\section{Experimental}
\label{sec:exp}

\begin{figure}
\centerline{\epsfig{file=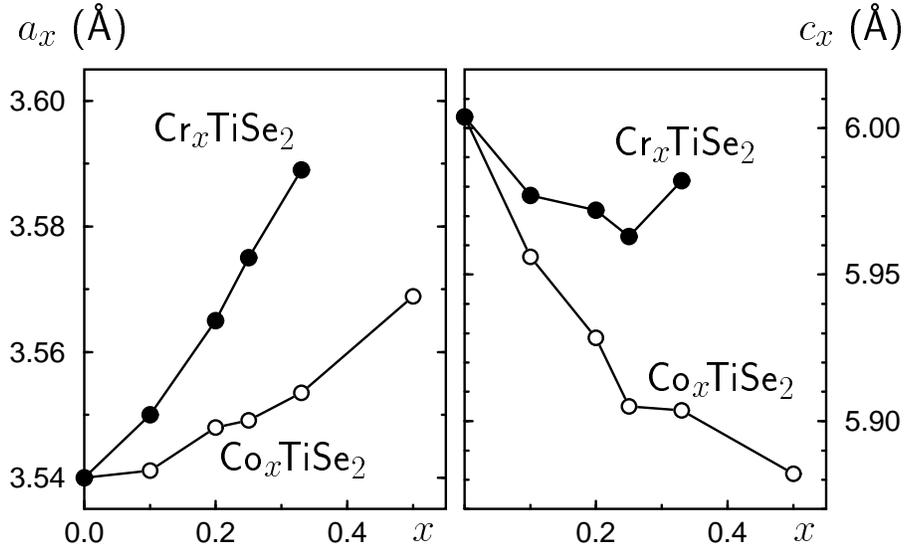,width=12.0cm}}
\bigskip
\caption{
The hexagonal unit cell parameters $a$ and $c$
of Co$_x$TiSe$_2$ and Cr$_x$TiSe$_2$ depending on $x$.
}
\label{fig:A+C}
\end{figure}

Polycrystalline Co$_x$TiSe$_2$ and Cr$_x$TiSe$_2$ samples ($0<x<0.5$) 
were prepared
by an ampoule synthesis method from the constituent elements. 
The methods of synthesis and characterization are described 
in detail in Ref.~\cite{FTT39-1618}. 
The initial diffraction data were obtained with a DRON 1UM diffractometer 
(Cu $K\alpha$, Ni filtered radiation.) 
The unit cell parameters have been then determined by a least square
refinement with 16 reflexes 
with the accuracy $\Delta a_0={\pm}0.001$, $\Delta c_0={\pm}0.002$ {\AA}.
The hexagonal unit cell parameters of intercalated TiSe$_2$
are shown in Fig.~\ref{fig:A+C}. 
The crystal structure, electric and magnetic properties 
of Co$_x$TiSe$_2$ are described in Refs.~\cite{JSSC17-9,FTT39-1618}. 
The information about the type of intercalant ordering in Co$_x$TiSe$_2$
is controversial \cite{JSSC17-9}, which may be traced back to different thermal 
treatments of the samples. In the present work we used 
the samples quenched from 850 $^{\circ}$C in ice water. 
For the samples so prepared, no ordering of intercalant atoms was 
observed in both systems. 

At temperatures above that of liquid nitrogen,
both materials are paramagnetic, with a temperature dependence 
of their magnetic susceptibility following a Curie-Weiss law.
The values of effective magnetic moments are close to nominal
spin moments of Co$^{2+}$ and Cr$^{3+}$ ions, respectively.

The single crystals for the X-ray spectra measurements were grown 
by the gas transport reaction method in evacuated quartz ampoules 
with I$_2$ as gas-carrier \cite{FMM81-629}. Powdered Co$_{0.25}$TiSe$_2$ 
and Cr$_{0.25}$TiSe$_2$ material was used as origin materials. 
In the case of Co$_{0.25}$TiSe$_2$ there was a flow of substance
from the hot (1000~$^{\circ}$C) to the cold (600~$^{\circ}$C) edge 
of the ampoule. The single crystals had a shape of thin plates 
$\approx$ 2-3 mm large and $\approx$ 0,1 mm thick. 
The final intercalant content (determined from the lattice 
parameter values) was found to be close to the initial one: 
Co$_{0.23}$TiSe$_2$. 
In the case of Cr$_{0.25}$TiSe$_2$ the substance was transferred
from the middle part (at 900~$^{\circ}$C) where the powder was initially 
placed to both hot (1100~$^{\circ}$C) and cold (700~$^{\circ}$C)
edges of the ampoule.
The single crystals grown at the hot edge of the ampoule 
had a shape of slim needles with the size close to
$0.3{\times}0.1{\times}10$ mm, whereas those grown at the cold edge
were thin plates 2--4 mm large and $\approx$0.1~mm thick.
From the unit cell parameters comparison we found that the crystals formed 
at the cold edge have the composition Cr$_{0.1}$TiSe$_2$, 
those at the hot edge -- Cr$_{0.5}$TiSe$_2$.

The X-ray photoelectron spectroscopy (XPS) measurements 
were performed at a Perkin Elmer spectrometer
with an energy resolution of 0.3 eV, 
X-ray emission spectroscopy (XES) measurements -- at a soft X-ray radiation 
spectrometer RSM-500 with electron excitation and an energy resolution
of 0.7 eV. The single crystals of Co$_{0.23}$TiSe$_2$ and
Cr$_{0.1}$TiSe$_2$ were cut by a razor blade in the spectrometer chamber
directly before the measurements.

\section{Crystal structure and bulk properties}
\label{sec:cryst}

\begin{figure}
\centerline{\epsfig{file=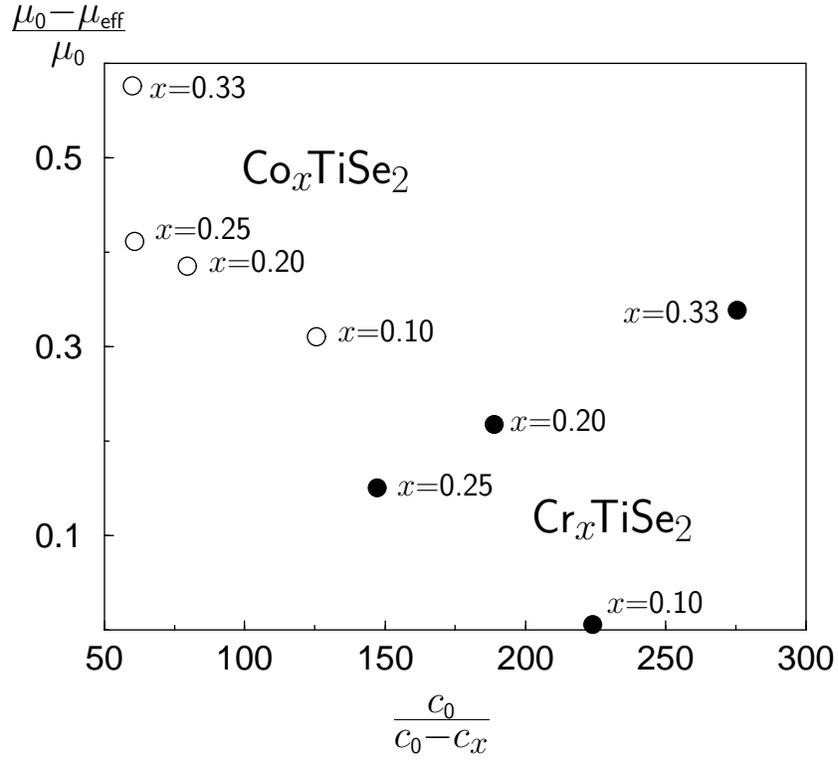,width=11.0cm}}
\bigskip
\caption{
Correlation between reduced magnetic moment of intercalant atom
$\mu_0-\mu_{\mbox{\tiny eff}}/\mu_0$
($\mu_0$ is the nominal free ion magnetic moment,
$\mu_{\mbox{\tiny eff}}$ as determined from magnetic susceptibility
measurements) and the lattice deformation parameter
$c_0/(c_0-c_x)$.
}
\label{fig:DmDc}
\end{figure}

In this section, we present the discussion of the electronic characteristics
as they might be deduced from the analysis of the measured bulk
properties. The microscopic aspects of hybridization
will be discussed in Sec.~\ref{sec:calcul} in relation with
{\em ab initio\/} calculations.
A decrease of the lattice parameter $c$ in the course of intercalation
is typical for $M_x$TiSe$_2$ systems
as has been mentioned in Ref.~\cite{JSSC17-9,FMM81-629} and could be seen in 
Fig.~\ref{fig:A+C}. This might be understood as a result of 
covalent bonds being formed between intercalant atoms 
and the host lattice \cite{Titov_MCLC98}. While free Co$^{2+}$ and 
Cr$^{3+}$ ions have the same spin value of 3/2, 
in intercalated systems
the experimentally obtained effective magnetic moment at the Cr atom 
is roughly 1.5 times larger than that per Co atom (see Table~\ref{tab:mom}). 
This fact may be related to a stronger hybridization of $3d_{xz,yz}$ orbitals 
(those participating in the magnetic moment formation 
in the $3d^7$ configuration of Co$^{2+}$) with the Se$4p$ orbitals.
In contrast, the $3d_{xz,yz}$ orbitals 
are essentially empty in the $3d^3$ configuration of Cr$^{3+}$.
The deviation of the effective moment from its nominal free-ion value
correlates well with the lattice distortion
(which is the measure of the
chemical bonding between an intercalant and the host \cite{Titov_MCLC98})
at least for concentrations $x{\leq}0.25$ as is shown in Fig.~\ref{fig:DmDc}. 
The systems with $x=0.33$ are actually beyond the percolation
limit on the ordered lattice of $3d$ ``impurities'', so that
their clustering and direct exchange interaction
apparently results in the drop of magnetic moments and breakdown
of the abovementioned correlation.

\begin{figure}
\centerline{\epsfig{file=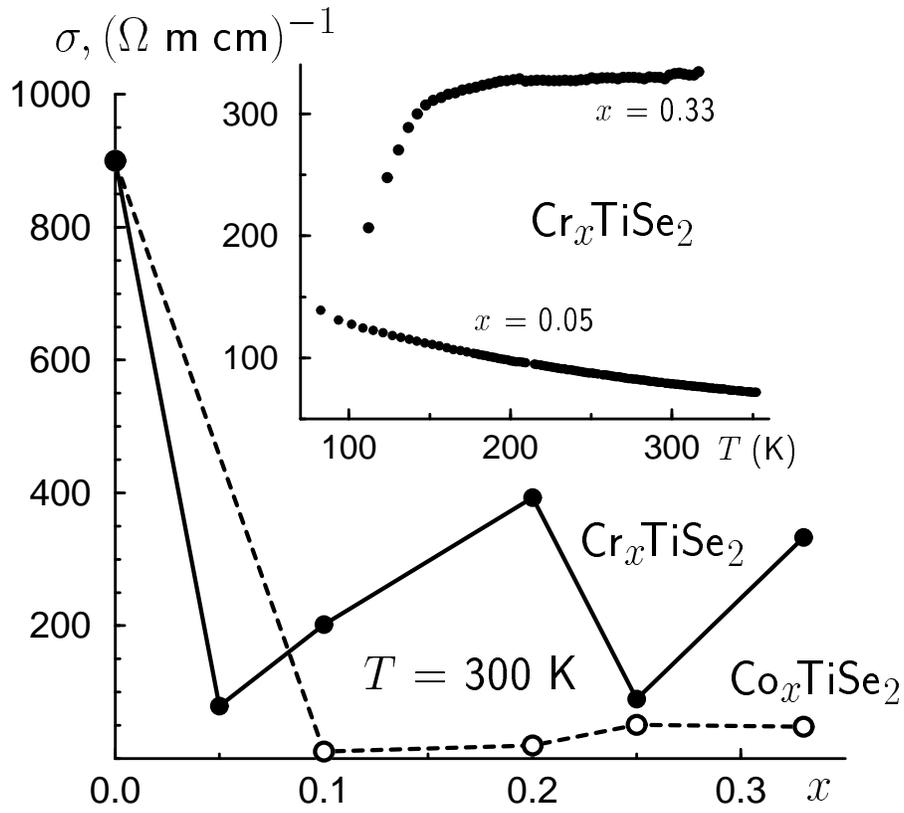,width=12.0cm}}
\bigskip
\caption{
The conductivity of Co$_x$TiSe$_2$ and Cr$_x$TiSe$_2$ as function 
of intercalant content at $T$=300~K. 
Inset: the temperature dependence of conductivity for 
Cr$_x$TiSe$_2$ with $x$=0.05 and 0.33.
}
\label{fig:Conduc}
\end{figure}

The lattice deformation and change in the degree of hybridization
affects the conductivity $\sigma$ of the $M_x$TiSe$_2$ systems. 
While in Co$_x$TiSe$_2$ $\sigma$ is only weakly concentration-dependent,
in Cr$_x$TiSe$_2$ it exhibits an interesting behavior --
being on the average higher, $\sigma(x)$ includes several irregularities.
Fig.~\ref{fig:Conduc} shows the conductivity at room
temperature, but the irregularities persist over a broad temperature
range. Two different types of $\sigma(T)$ are shown in the inset
of Fig.~\ref{fig:Conduc} for two Cr concentrations.

An overall decrease of $\sigma$ as the unit cell parameter $c$ decreases 
(see Fig.~\ref{fig:Conduc}) is apparently due to the drop
in the free charge carriers concentration as electrons are
trapped in covalent bonds -- this seems to be well the case
at least in all Co-containing systems. Due to a different charge configuration
of Cr$^{3+}$ as compared to Co$^{2+}$,
one electron provided by Cr remains free and contributes
to the metallic type of conductivity --
see the calculated density of states (DOS) distribution
at the impurity site, Sec.~\ref{sec:calcul}. 
A drop in the conductivity near $x$=0.05 of Cr
probably indicates the compensation 
of highly mobile intrinsic (hole-type) carriers in TiSe$_2$,
the concentration of which is 0.05 per formula unit\cite{PSSb86-11}. 
This is consistent with a perfectly metallic type of temperature
dependency of conductivity for this concentration,
as shown in the inset of Fig.~\ref{fig:Conduc}.
Another drop in $\sigma(x)$ at $x$=0.25 may be related to the percolation limit 
on a lattice of those Ti atoms which have no Cr neighbors.
Two Ti atoms (to both sides of the intercalated Cr atom)
participate in the formation of localized bonds,
and the percolation limit for two-dimensional hexagonal lattices 
equals 0.5 \cite{JMP5-1117}. 
For $x$=0.33, differently from lower Cr concentrations,
the activation region is seen (inset in Fig.~\ref{fig:Conduc})
in the $\sigma(T)$, indicating an opening of the band gap at the Fermi level.
It is noteworthy that at $x$=0.33,
exactly one extra electron is provided by Cr per titanium atom.
One can speculate that the gap formation is the consequence 
of spin splitting of the Cr-Ti hybridized band, 
since this band could exactly accommodate one electron per Ti atom. 
No such gap is seemingly formed in Co$_x$TiSe$_2$,
presumably because more efficient deformation of the host lattice 
results in more octahedral coordination of Ti atoms by chalcogen, 
preventing the splitting between  
$d_{z^2}$ and $d_{xy}$, $d_{x^2-y^2}$ orbitals of Ti \cite{Liang_NATO84}.

\section{X-ray photoelectron and X-ray emission spectra}
\label{sec:xps}

\begin{figure}
\centerline{\epsfig{file=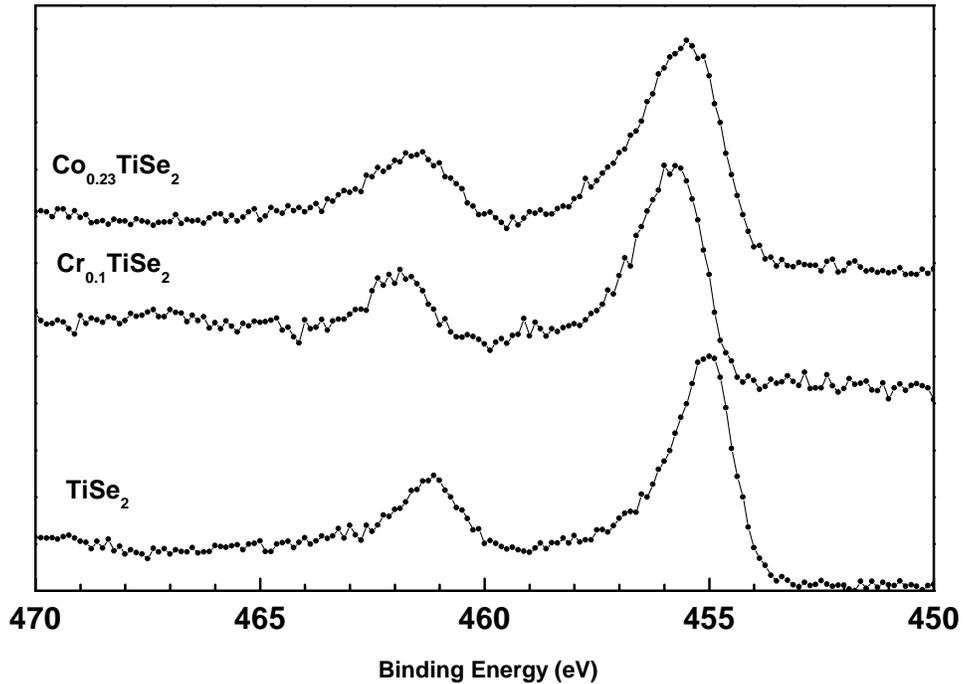,width=13.0cm}}
\bigskip
\caption{
X-ray photoelectron spectra of Ti $2p$ core levels 
in Co$_{0.23}$TiSe$_2$, Cr$_{0.1}$TiSe$_2$, and TiSe$_2$. 
}
\label{fig:Ti2p}
\end{figure}

\begin{figure}
\centerline{\epsfig{file=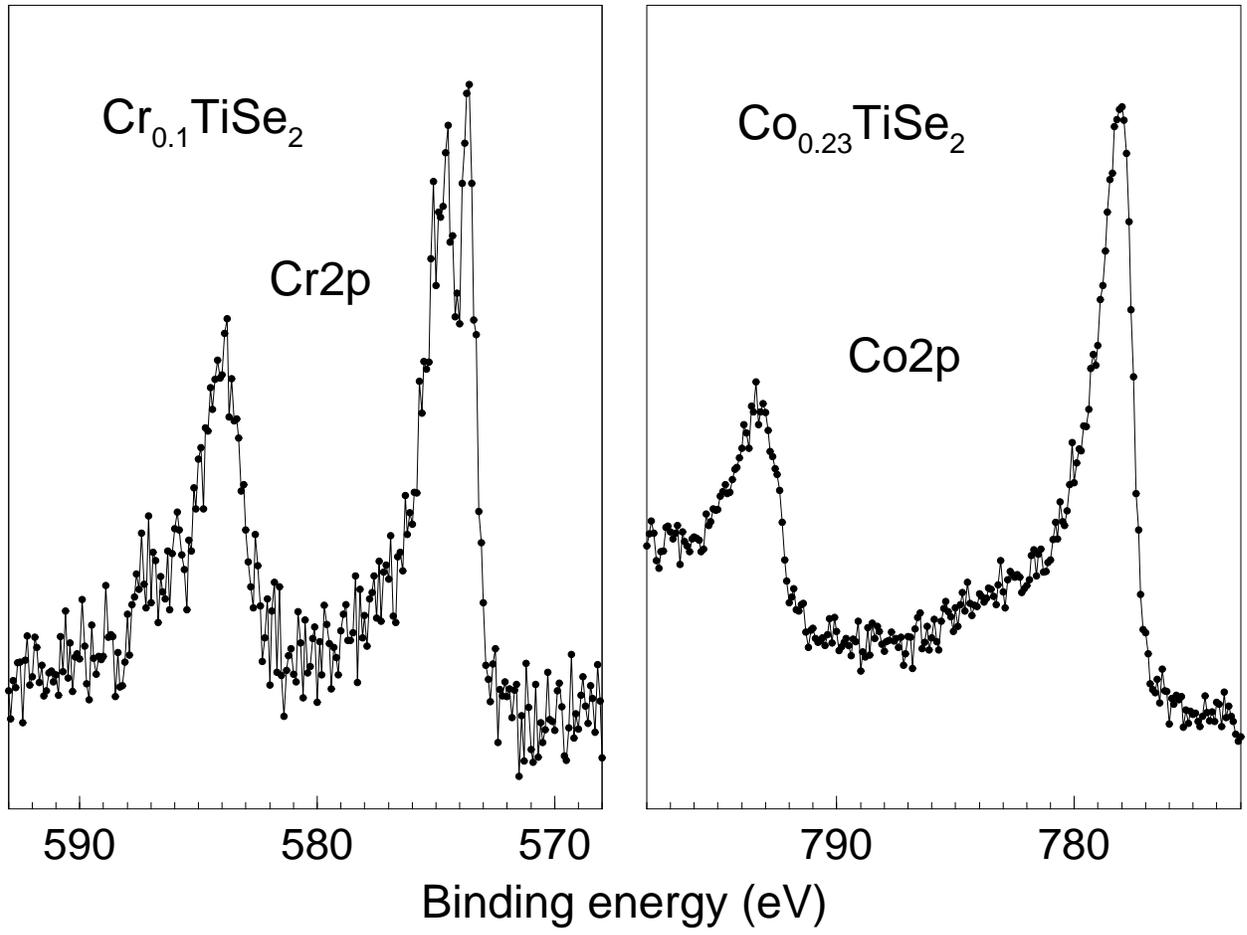,width=16.0cm}}
\bigskip
\caption{
X-ray photoelectron spectra of intercalant
$2p$ core levels in Cr$_{0.1}$TiSe$_2$ and Co$_{0.23}$TiSe$_2$.
}
\label{fig:XPScore}
\end{figure}

We refer to the XPS spectra with three main purposes: to control
the quality of monocrystalline samples, to get information about
the electronic structure and to trace the effects of on-site magnetic
interactions on the spectra.
Figs.~\ref{fig:Ti2p} and \ref{fig:XPScore} depict the $2p$ core levels of
titanium and of intercalated elements, correspondingly.
Ti $2p$ spectra are known to exhibit relatively large
chemical shifts already at early oxidation states 
(see, e.g., \cite{XPS_Handbook}). The actual spectra indicate
a good (oxygen-free) quality of samples.
The chemical shift still present 
(455.0 eV for the Ti$2p_{3/2}$ binding energy in  TiSe$_2$;
455.5 eV in Co$_{0.5}$TiSe$_2$; 455.8 eV in Cr$_{0.1}$TiSe$_2$)
exhibits the increase of ionicity from the host
to Co-based to Cr-based systems \cite{comment_MTiS2}.
This is consistent with above discussed (Fig.~\ref{fig:A+C})
stronger $z$-contraction of the crystal lattice 
in more covalent Co$_x$TiSe$_2$, as compared to Cr$_x$TiSe$_2$.

Of interest are the core $2p$ photoemission spectra of intercalant
atoms, which may exhibit an exchange splitting due to an
intra-atomic $2p$--$3d$ interaction (similarly to what has been demonstrated
for the Mn-based Heusler alloys\cite{Heus_EPJ98,PRB60-6428}). 
As is seen in Fig.~\ref{fig:XPScore}, the Cr$2p_{3/2}$ line
in Cr$_{0.1}$TiSe$_2$ clearly shows a splitting of about 1 eV,
differently from the Co$2p_{3/2}$ spectrum of Co$_{0.23}$TiSe$_2$
where no such splitting is observed. The splitting is qualitatively
related to the difference between local magnetic moments in 
these compounds (see Table \ref{tab:mom}).

\begin{figure}
\centerline{\epsfig{file=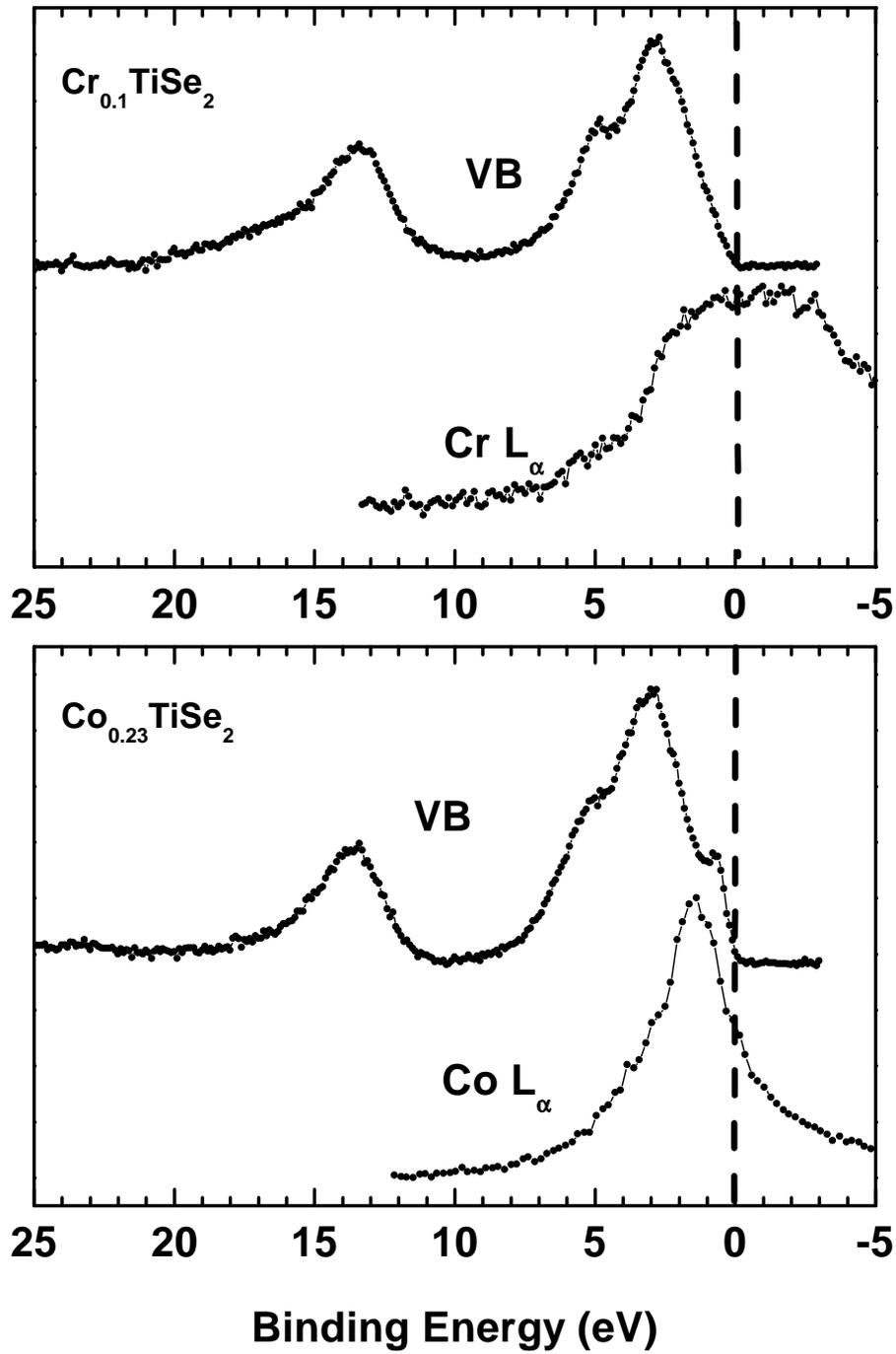,width=12.0cm}}
\bigskip
\caption{
X-ray photoelectron spectrum of the valence band (VB)
and $L_{\alpha}$ X-ray emissioum, brought
into the same scale of binding energies for 
Co$_{0.1}$TiSe$_2$ (top panel) and
Cr$_{0.23}$TiSe$_2$ (bottom panel). Note the contribution
to the Cr$L_{\alpha}$ spectrum from the states above the Fermi energy.
}
\label{fig:XPS+XES}
\end{figure}

The valence-band (VB) photoelectron spectra 
(Fig.~\ref{fig:XPS+XES}) have a relatively
simple structure, where the Se$4s$ contribution
(in the region of binding energies 12--16 eV) is clearly separated
from the Se$4p$--Ti$3d$--$M$$3d$ hybridized band
spanning the binding energies 0--6 eV. 
In the latter band, the contribution from the
Se$4p$ states is the most pronounced, since the photoionization
cross-section of selenium is much larger than for all other
constituents \cite{YeLin_ADNDT85}. However, the contribution of
Cr$3d$ and, in particular, Co$3d$ states is noticeable
just below the Fermi energy, similarly to the behavior
detected for the Fe$3d$ states in Fe$_x$TiTe$_2$ \cite{Titov_MCLC98}.
An analysis of the calculated DOS (see next section) shows that 
a narrow peak at $\approx$1.5 eV below the Fermi level,
well separated from the bulk of Se$4p$ states, indeed
dominates the local DOS at the Co site and 
must be visible in the valence-band XPS.

\begin{figure}
\centerline{\epsfig{file=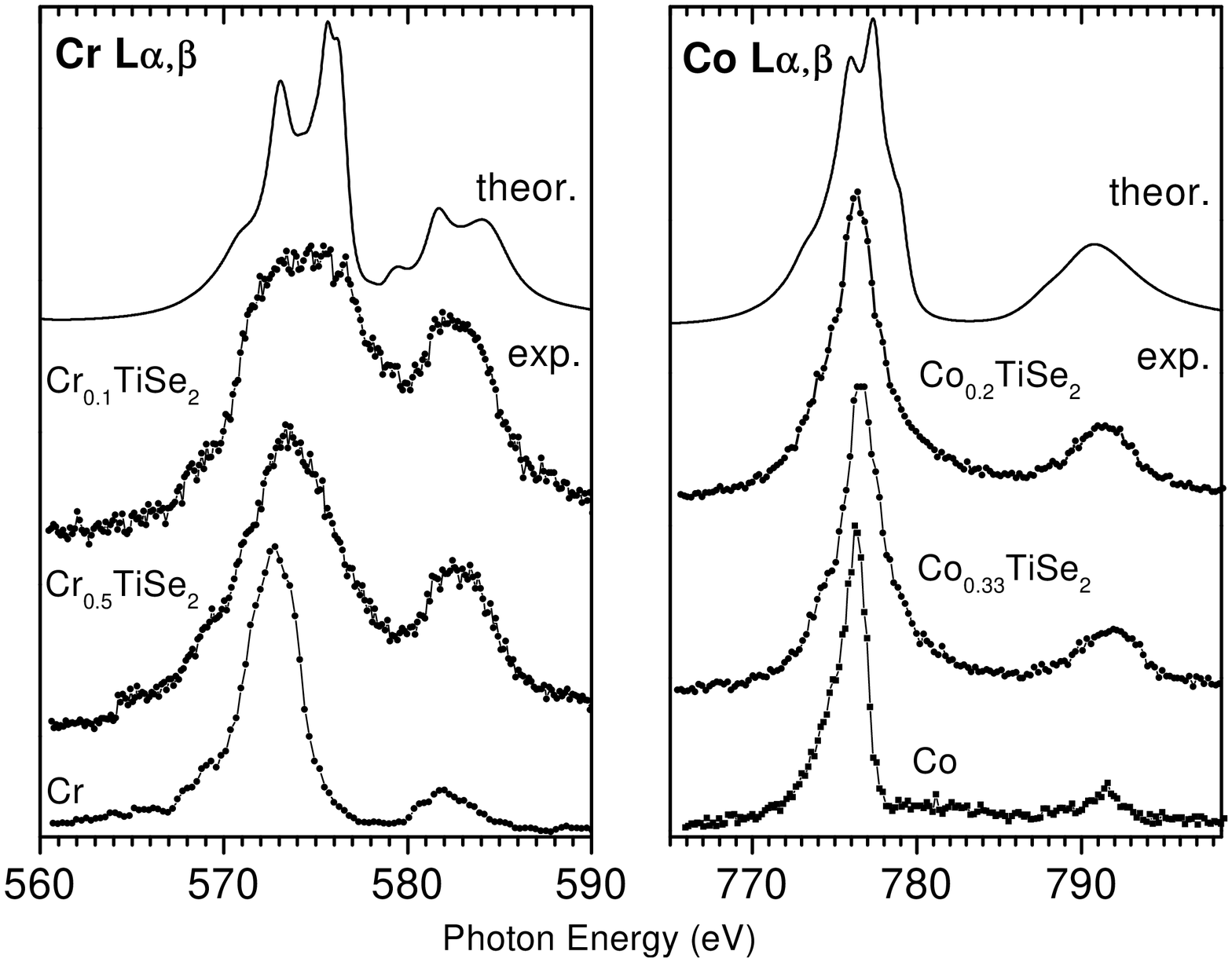,width=12.0cm}}
\bigskip
\caption{
$L_{\alpha,\beta}$ X-ray emission spectra of Cr and Co
in pure metals (bottom) and intercalated TiSe$_2$.
Top curve: the spectra calculated from first principles
for the $M$(TiSe$_2$)$_3$ ($M$=Cr, Co) supercell;
see text for details.
}
\label{fig:XES}
\end{figure}

The trends in the ``local'' electronic structure of intercalant atoms
depending on concentration are much better visible
from X-ray emission spectra which are element-selective
and hence not masked by the contributions from
selenium states. The Cr and Co $L_{\alpha,\beta}$ spectra
($4s3d \rightarrow 2p_{3/2,1/2}$ transitions)
have been measured for several concentrations
and are shown in Fig.~\ref{fig:XES}.
Based on the arguments of Ref.~\onlinecite{FTT38-1709},
the concentration-related differences in the spectra
can be traced to the percolation
threshold on a two-dimensional hexagonal lattice at $x=0.25$.
Above this threshold, the intercalant atoms (under an uniform
distribution) may have a neighbor of the same type
at the distance of hexagonal lattice constant.
It is noteworthy how the width and the relative
intensity of $L_{\alpha}$ and $L_{\beta}$ bands vary
with the concentration. The Cr $L_{\alpha,\beta}$ spectra
of intercalated compounds are quite different from those
of pure metallic Cr, whereas the corresponding differences
for the Co-based systems are less pronounced. 

A striking difference in the XES
of Co- and Cr-intercalated systems is a much broader emission band
in the latter, in spite of the fact that the widths of the occupied
valence bands in both systems are presumably not too much different. 
The shift of the X-ray emission spectra 
by the $2p_{3/2}$ core level binding energy 
onto the common energy scale with the XPS (Fig.~\ref{fig:XPS+XES})
makes obvious that the states above the nominal Fermi level 
effectively participate in the emission process in Cr$_{0.1}$TiSe$_2$, 
but not in Co-intercalated systems.  
Another peculiarity of the spectra shown in Fig.~\ref{fig:XES}
is the increase of the $L_{\beta}/L_{\alpha}$ relative intensity
in more diluted (especially in Cr-based) systems.
The broadening of $L_{\alpha}$ spectra 
of $3d$ components has been detected and discussed before -- 
in $3d$ oxides (of Mn \cite{PRB54-4405,PRB55-4242} 
and Cu \cite{Kawai_Cu_92}) where charge-transfer excitations were argued 
to be an important mechanism responsible for a rich satellite structure, 
as well as in pure $3d$ metals (Cu, \cite{PRB56-12238}) where an initial-state 
$3d$ vacancy satellite was shown to contribute on the high-energy side of the
$L_{\alpha}$ line for excitation energies beyond the $L_2$
threshold. For our systems in question, none of these
mechanisms seem to work: 
charger-transfer excitations cannot play an important
role because of a much high covalency than in oxides;
as for shake-up/shake-off processes -- it is difficult to argue
why their intensity would be enhanced at low Cr (but not Co)
concentrations. Moreover, the increase of the high-energy satellite
in Cu $L_{\alpha,\beta}$ in Ref.~\cite{PRB56-12238}
is accompanied by the decrease of the $L_{\beta}/L_{\alpha}$
intensity ratio (due to the Coster-Kronig process) -- the trend
opposite to that observed in the present case.

What makes the difference between low-concentrartion
Cr-intercalated system and other systems studied
is relatively high localization of the Cr$3d$ state and
high magnetic moment associated with it. It is noteworthy that
in recently studied Heusler alloys NiMnSb and Co$_2$MnSb
where the high magnetic moment at the Mn site is 
similarly localized, the $L_{\alpha}$ emission from 
the states nominally above the Fermi level was observed
\cite{Heus_dichro}.
Similarly to how it was argued in Ref.~\cite{Heus_dichro}
for Heusler alloys, we tend to qualitatively attribute the anomalous 
X-ray emission in the diluted Cr$_x$Tise$_2$ system to 
a relaxation of the intermediate state with electronic configuration
$2p^53d^{n+1}$, which involves the contribution from the states
of $d$ symmetry nominally vacant in the ground-state configuration
$2p^63d^n$. Below, we shall refer to this effect for brevity
as re-emission. Strictly speaking, the excitation and subsequent emission
in a resonant process must be treated as a single
event; an extensive literature exists on this subject
(see, e.g., Refs.~\onlinecite{PRB94-5799,APA65-91,APA65-159}).
The effect is the more pronounced the stronger the
localization of involved states is. The peculiarity of our present
spectra is that such typically resonance-like phenomenon
as re-emission is observed in spite of high-energy 
(i.e. non-resonance) electronic 
excitation, with the excitation energy of nearly 4 KeV,
apparently due to sufficiently high lifetime of an electron
trapped in a highly localized vacant Cr $3d$ state.
For the Co-intercalated systems, the effect is practically
not observable.
Assuming phenomenologically the participation of the lowest vacant 
Cr $3d$ states in the X-ray emission, we leave the (otherwise single-particle)
quantitative treatment of the spectrum
till next section where the microscopic information about  
the electronic structure becomes available.

Another important observation concerning the X-ray emission
spectra is a substantial increase -- especially at low 
intercalant concentrations --  
of the $L_{\beta}/L_{\alpha}$
intensity ratio as compared to that in the pure metal.
To our opinion, this indicates a localization
of the $3d$ electron density and hence a magnetic moment.
$L_{\alpha}$ and $L_{\beta}$ bands originate from the dipole transitions
$2p_{3/2}{\rightarrow}3d_{5/2,3/2}$ and
$2p_{1/2}{\rightarrow}3d_{3/2}$, correspondingly. 
The change in the $L_{\beta}/L_{\alpha}$ relative intensity may be influenced
by the selection rule on the total moment projection, ${\Delta}m_j=0$, 
which in the present case favors a substantial decrease 
of the statistical weight of the $2p_{3/2}{\rightarrow}3d_{5/2,3/2}$ transition,
since the exchange interaction energy of $3d$ electrons
(2--2.5 eV) by more than an order of magnitude exceeds 
their spin-orbit interaction energy. 

As already mentioned above, the changes in
the observed $L_{\beta}/L_{\alpha}$ weights distribution
for systems with lower Cr content is opposite to what one could
expect from the Coster-Kronig transitions, which redistribute
intensities from $2p_{1/2}$ to the $2p_{3/2}$ channel.
This indicates a relative unrelevance
of the Coster-Kronig effect for the systems in question.

\section{Electronic structure calculations}
\label{sec:calcul}

In Ref.~\onlinecite{EMRS99} we reported the results
of a first-principles optimization of lattice parameters in
pure TiSe$_2$ (as well as in TiTe$_2$) and the electronic
structure of $3d$-intercalated systems, with the use of
$M$(TiSe$_2$)$_3$ supercells in the latter case. 
The calculation has been done with the full-potential linearized
augmented plane wave code WIEN97\cite{wien97}, using the
generalized gradient approximation (GGA) in the formulation
by Perdew, Burke and Ernzerhof\cite{GGA_PBE_96}.
The hexagonal lattice parameters as found in the course
of total energy minimization were $a$=3.519 {\AA},
$c$=6.280 {\AA}, whereas the experimental measurements
give the values $a$=3.535 {\AA}, $c$=6.004 {\AA}
(as cited in Ref.~\cite{PRB51-17965}).
The internal coordinate $z$ (the relative distance between Ti 
and Se layers) is 0.247 in the calculation, compared
to the measured value of 0.25.
At a first sight, the GGA overestimates the $c/a$ ratio
by as much as 5\%. However, considering the reduced value of $z$
in the calculation, one can notice that in reality just
the van der Waals gap between adjacent chalcogen planes is overestimated,
whereas the lenghts of Ti-Ti and Ti-Se bonds remain well within 1\% of
the experimental estimate. 
The Perdew--Burke--Ernzerhof formula for gradient approximation
was optimally tuned for a certain degree of localization 
of the exchange-correlation hole, typical for bulk solids
but not for the ``empty'' van der Waals gap. Some discussion
on this subject can be found, e.g., in Ref.~\cite{TSRNC_98}.
In the present study, we consider systems with relatively high
concentration ($1/3$ per TiSe$_2$ unit) of intercalant atoms
in the van der Waals gap, therefore the abovementioned shortcoming
of the GGA treatment must be less serious than for pure
dichalcogenides.

\begin{figure}
\centerline{\epsfig{file=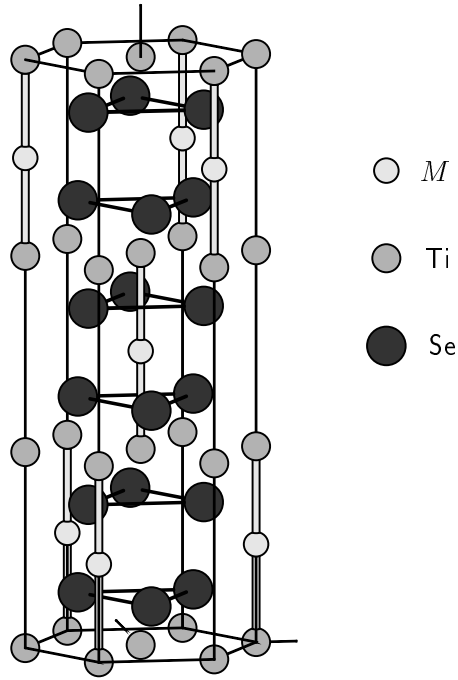,width=6.0cm}}
\bigskip
\caption{
The $M$(TiSe$_2$)$_3$ supercell (in the hexagonal setting) used 
in the {\em ab initio\/} electronic structure calculations ($M$=Cr, Co).
}
\label{fig:Scell}
\end{figure}

As in Ref.~\cite{EMRS99}, we simulated intercalated systems using the 
rhombohedral $M$(TiSe$_2$)$_3$ supercell,
shown in Fig.~\ref{fig:Scell} (in the hexagonal setting).
Similar supercells were earlier used in non-magnetic calculations
by Suzuki {\em et al.\/}\cite{SYM_JMMM87,SYM_JPSJ89}. 
With only one type of supercell being considered in {\em ab initio\/}
calculations, it had little sense to address the systematical trends in the 
lattice parameters under doping. Instead, we concentrated
in our analysis on the properties of two intercalated systems
as {\em impurity\/} systems. The calculations have been performed
therefore in a fixed ($c$, $a$) frame, with a local atomic
relaxation around the $M$ atoms taken into account. This allows us to make
a more clear comparison of Cr- and Co-intercalated systems.
From the following it becomes obvious that the noticeably perturbed region
around an intercalant atom is indeed much smaller than the region defined
by the translation vectors in the supercell of our choice.

In the hexagonal setting, the supercell with the lattice constants
$a'=a\sqrt{3}$, $c'=3c$ (Fig.~\ref{fig:Scell})
hosts three $M$(TiSe$_2$)$_3$ units
with the following atomic positions in one of them: $M$ at (0,0,0);
Ti at $(0,0,\frac{1}{2})$ and at 
$\pm(0,0,\frac{1}{6}\!+\!\Delta_1)$;
six Se at 
$\pm(1\!-\!\Delta_2\!-\!\Delta_3,
\frac{1}{3}\!-\!\Delta_3,
\frac{5}{12}\!+\!\Delta_4)$;
$\pm(\frac{2}{3}\!+\!\Delta_3,
\frac{2}{3}\!-\!\Delta_2,
\frac{5}{12}\!+\!\Delta_4)$;
$\pm(\frac{1}{3}\!+\!\Delta_2,
\Delta_2\!+\!\Delta_3,
\frac{5}{12}\!+\!\Delta_4)$.
The optimized values of internal coordinates $\Delta_1$ to $\Delta_4$,
obtained from the condition that the forces on all atoms disappear,
are listed in Table \ref{tab:relax}. The local structural distortion,
as compared to the equilibrium structure of TiSe$_2$, is quite small.
The negative value of $\Delta_4$ corresponds to the displacement
of the chalcogen planes towards the Ti layers by 0.42\% of 
the $c$ parameter of TiSe$_2$; this corresponds roughly
to $z=0.247$ rather than 0.25 found for pure TiSe$_2$. 
A qualitative difference between Cr$_{1/3}$TiSe$_2$ and
Co$_{1/3}$TiSe$_2$ is that the nearest Ti atoms approach 
the Co atom ($\Delta_1$ slightly negative) but are shifted away
from the Cr atom. This is consistent with the evidence of
larger Co--Ti bonding and, contrarily, higher localization 
of the Cr$3d$ states as addressed below.

\begin{figure}
\centerline{\epsfig{file=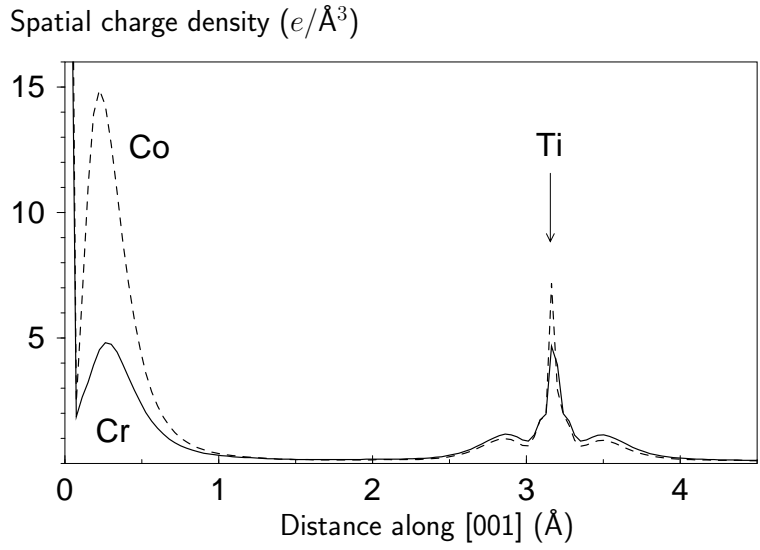,width=10.0cm}}
\bigskip
\caption{
The cut of the charge density distrubution along the [001]
from the $M$ position in $M$(TiSe$_2$)$_3$ ($M$ = Cr, Co).
}
\label{fig:Dens_line}
\end{figure}

\begin{figure}
\centerline{\epsfig{file=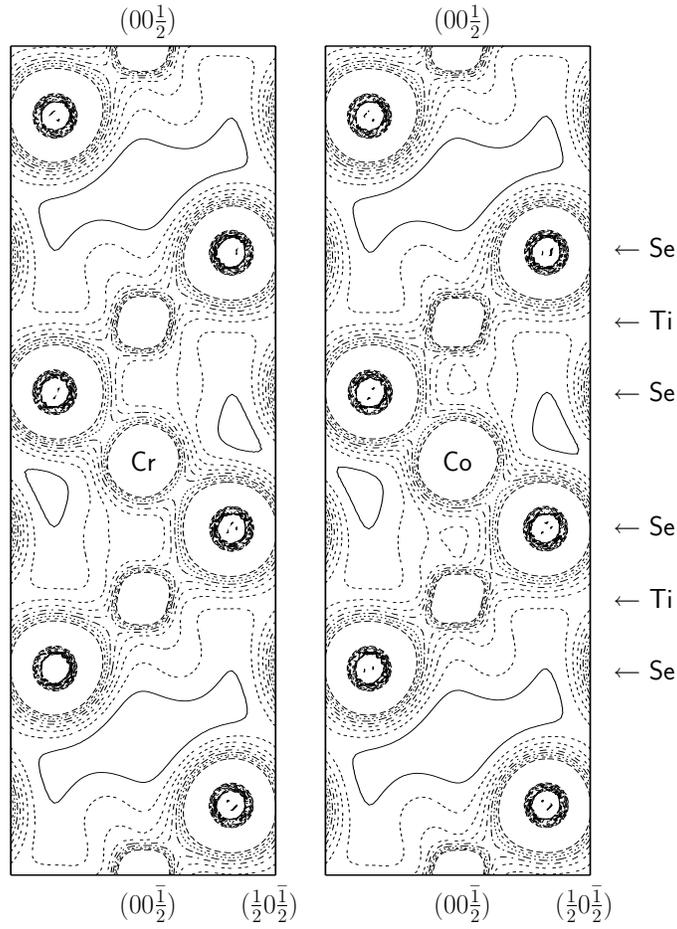,width=9.0cm}}
\medskip
\caption{
The contour plot of the charge density distrubution
over the supercell. The points labeled are in the 
supercell hexagonal setting ($c'=3c$, $a'=\sqrt{3}a$ -- see text).
The contours go up to 0.5 $e$/{\AA}$^3$ with the step 
0.05 $e$/{\AA}$^3$.
}
\label{fig:Dens_cont}
\end{figure}

\begin{figure}
\centerline{\epsfig{file=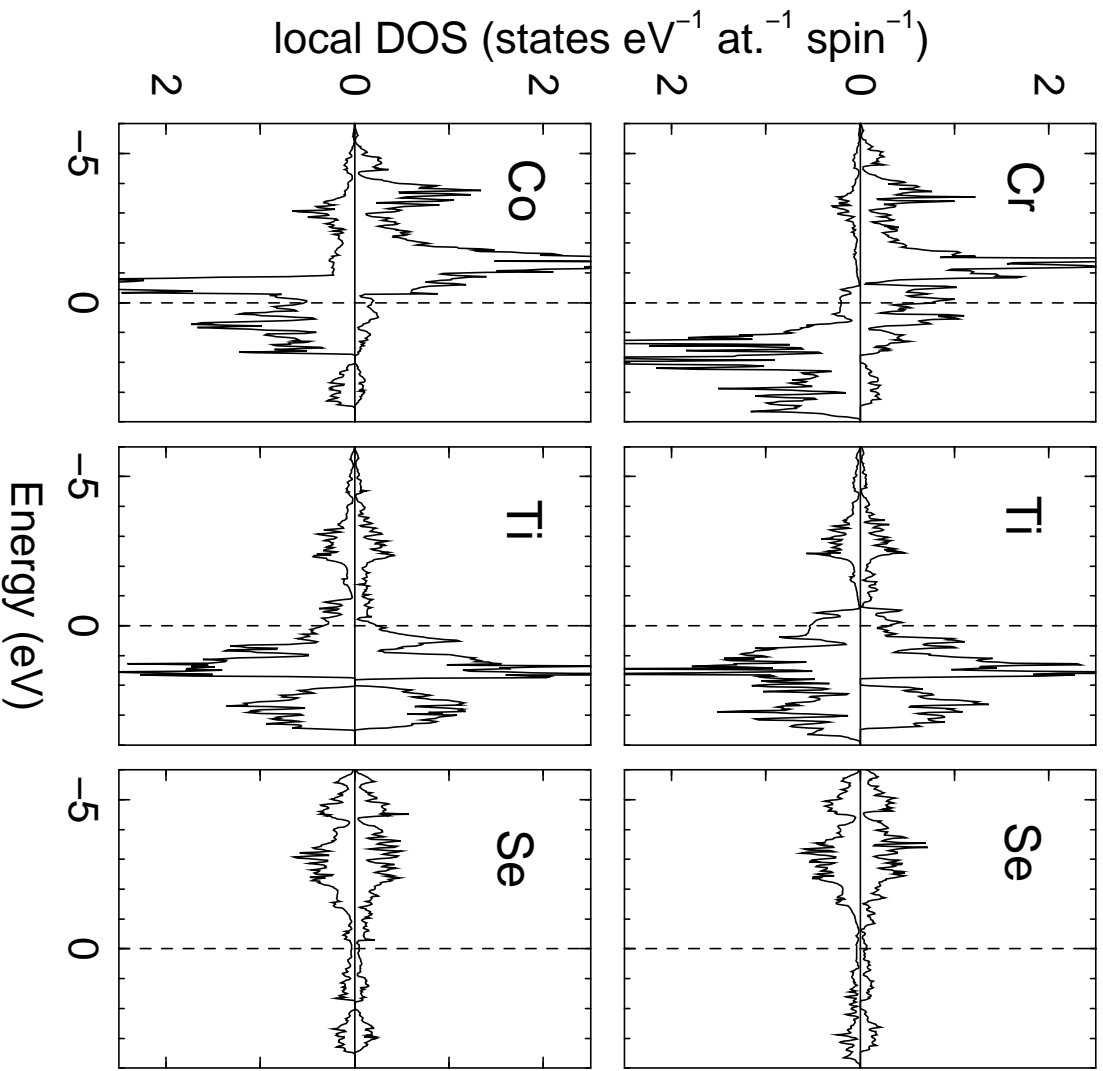,width=14.0cm}}
\bigskip
\caption{
$M3d$, Ti$3d$ (at the Ti atoms closest to $M$),
and Se$4p$-DOS as calculated for the $M$(TiSe$_2$)$_3$ supercell.
Upper row: $M$ = Cr, lower row: $M$ = Co.
}
\label{fig:DOS}
\end{figure}

The spatial distribution of the charge density in a plane cut 
across the supercell is shown in Fig.~\ref{fig:Dens_line}.
The vertical size of the plot covers 18.84 {\AA}. One can see
that in the absence of an $M$ atom the charge density
between the Se layers is indeed rather low, about
0.05 $e$/{\AA}$^3$. An intercalant atom embedded
in this region mediates the chemical bonding between
chalcogen atoms. An effectively larger (see below) Co atom 
provides a slightly higher charge density along the Se--$M$ line.
Moreover, the distrubution of the charge density
along [001] (i.e., towards the nearest Ti atoms) is somehow
different in Cr- and Co-intercalated crystals (Fig.~\ref{fig:Dens_cont}). 
On the Co site, the occupation of $3d$ states is substantially
higher as compared to Cr, with the effect that the effective size
of an intercalant atom is larger, in spite of a slightly higher
contraction of the Co$3d$ states.
At the same time, the occupation of the Ti$3d$ shell is lower
in the vicinity of Co next to a Cr atom. The local DOS
distribution on the atoms in question 
(Fig.~\ref{fig:DOS}) helps to understand the origins and energetics 
of this different charge distribution. 

The center of gravity of the Cr$3d$ states lies closer to that
of Ti$3d$. In spite of large exchange splitting at the Cr site,
a noticeable contribution from both atoms is present
near the Fermi level. Specifically, mostly the
$3d_{z^2}$ orbitals at both Cr and Ti sites are responsible
for this direct hybridization and the enhancement of the Ti DOS
(of the atoms neighboring to Cr in the [001] direction)
near the Fermi energy. This factor is not important
in the Co-intercalated system because the Co$3d$ states are lower
in energy and do not mix so efficiently with Ti$3d$ -- in fact, 
the majority-spin band at the Co site is fully occupied. 
Apart from this enhancement of the direct Cr--Ti bonding,
the distribution of electronic density (spatial as well as
energy-resolved) within the TiSe$_2$ remains largely unaffected
by the presence of intercalant atoms
and very much close to that of pure TiSe$_2$ (see
Ref. \cite{YSKim_JJAP98}. We notice however a difference
of our electronic structure from that calculated for TiSe$_2$
in Ref.~\cite{PRB51-17965}, with the use of the muffin-tin
approximation).

The local DOS at the $M$ site contains spin-splitted
and broadened levels, typical for a magnetic impurity
embedded in the VB (the latter being formed, in this case,
by the Ti$3d$--Se$4p$ hybridization). An anomalously narrow
minority-spin peak appears at the Co site because
its energy position is between the separated subbands
due to a bonding and antibonding Ti--Se interaction
(see more detailed discussion in Ref.~\cite{YSKim_JJAP98}),
i.e. the scattering of valence-band electrons at this impurity
level is suppressed. A more detailed analysis of the crystal-field 
splitting and spatial distribution of the $M3d$ orbitals
will be given elsewhere \cite{Ovchin}.
It should be noted that the present calculation for the ordered
Cr$_{1/3}$TiSe$_2$ structure failed to find an energy gap
at the Fermi level (implied by the conductivity type
as shown in Fig.~\ref{fig:Conduc} for $x$=0.33). The reason may be either
an effect of the lattice parameters' change, important for
the Cr-intercalated systems but ignored in the present calculation,
or the importance of correlation effects due to high Cr$3d$ localization,
that would justify the use of a different (orbital-dependent)
form of the total energy functional. This subject yet needs
a further clarification. 

The values of local magnetic moments (attributed to
muffin-tin sphere sizes cited above) are 2.81 $\mu_{\mbox{\tiny B}}$
for Cr and 1.46 $\mu_{\mbox{\tiny B}}$ for Co, that agrees well
with the experimental values of Table \ref{tab:mom}
for $x$=0.33. In spite of a slight variation of the magnetic density
over the crystal (small antiparallel polarization of Ti$3d$ shells
and parallel of the Se$4p$ shells), the magnetic moment remains
quite localized at the intercalant atom. For the values of the $2p$ 
exchange splitting (taken simply as energy differences between
spin-up and spin-down components of self-consistently calculated
relativistic core levels), we obtained 0.85 eV at the Cr site
and 0.59 eV at the Co site. The former is in good agreement
with the experimental estimate, and the latter is probably
already too small to be resolved with the experimental resolution
(see Fig.~\ref{fig:XPScore}).

With the information about the local state density distribution
at intercalant sites available, it becomes possible to
address the problem of X-ray re-emission in the Cr$_x$TiSe$_2$
systems with low $x$ content. An assumption of a strongly localized 
character of Cr$3d$ states is supported by the calculations
of both spatial distribution (Fig.~\ref{fig:Dens_line}) and the local DOS  
(Fig.~\ref{fig:DOS}). Localized states in the vacant region
(which may host an absorbed electron in an intermediate state
of a X-ray emission process) must be necessarily those in the
minority-spin channel. 
In order to imitate their possible participation
in the X-ray re-emission, we calculated the Cr$L_{\alpha,\beta}$ 
X-ray emission
spectrum, taking into account dipole transition matrix elements
and the cutoff energy of 2 eV above the nominal Fermi level,
to allow the lowest minority-spin band
localized at the Cr site to contribute in full. 
This band incorporates more than two extra electrons,
rather than one as would be the case in a simple intermediate-state
model. However, there is no simple criterium to select
only part of these low-lying vacant states for their possible
contribution to the re-emission process. 
The result of a calculation is shown in
Fig.~\ref{fig:XES}, top curve. Without the Fermi energy 
artificially shifted, the higher peak in both $L_{\alpha}$ and
$L_{\beta}$ bands is missing, so that the calculated spectrum
is rather close to that of pure chromium. 
No such artificial adjustment was necessary for the Co$_x$TiSe$_2$
system.

\section*{Conclusion}

As a result of this combined study incorporating measurements of
the crystal structure, kinetic properties (conductivity) and
the analysis of the electronic structure by different spectroscopic
techniques (XPS and XES), interpreted on the basis of
electronic structrure calculations, we establish the following.
In spite of nominally identical magnetic moments of
a free ion for Cr and Co, the saturation magnetization 
of intercalated TiSe$_2$ reveals
an almost two times larger effective moment on Cr than that 
on Co sites. A large exchange splitting in the Cr$3d$ shell
polarizes the Cr$2p$ shell and leads to the exchange splitting
of Cr$2p$ core levels large enough (0.9 eV) to be detected
by photoelectron spectroscopy. No such splitting was
resloved for Co$2p$ levels.
So different magnetic behavior can be traced to different
strength of chemical bonding in Cr- and Co-intercalated
systems. The bonding is stronger in Co$_x$TiSe$_2$,
as is evidenced by larger distortion of the host lattice.
The Cr$3d$ states, on the contrary, remain more localized,
that favors a higher magnetic moment. Another indication
of higher localization of Cr$3d$ states in Cr$_x$TiSe$_2$
is a strong contribution in the X-ray emission $L_{\alpha,\beta}$ spectrum
from above the nominal Fermi level,
observed even at excitation energies much higher than the threshold 
and tentatively attributed to re-emission.  
We emphasize that the exchange splitting of $2p$ core levels
and a strong re-emission due to high-energy electronic excitations
are experimental findings quite rarely observed 
in $3d$-based systems so far. The reported first-principles
calculations provide an adequate description of the structural and
magnetic properties and simplify at least a qualitative
understanding of the underlying physics.
However, an attempt to take into account correlation effects
more systematically could be interesting for the treatment
of highly localized Cr$3d$ states and of the excitation processes
involving the latter.

\section*{Acknowledgment}

This work was supported by the Russian Science Foundation for Fundamental
Research (Projects 96-15-96598 and 99-03-32503), 
a NATO Linkage Grant (HTECH.LG 971222) and a DFG-RFFI Project.
Financial support from the Deutsche Forschungsgemeinschaft
is greatly acknowledged.

%\bibliographystyle{apsrev}
%\bibliography{chalco,spectro,methods,notes}

\begin{thebibliography}{10}
\expandafter\ifx\csname bibnamefont\endcsname\relax
  \def\bibnamefont#1{#1}\fi
\expandafter\ifx\csname bibfnamefont\endcsname\relax
  \def\bibfnamefont#1{#1}\fi
\expandafter\ifx\csname url\endcsname\relax
  \def\url#1{\texttt{#1}}\fi
\expandafter\ifx\csname urlprefix\endcsname\relax\def\urlprefix{URL }\fi
\expandafter\ifx\csname bibinfo\endcsname\relax \def\bibinfo#1#2{#2}\fi
\expandafter\ifx\csname eprint\endcsname\relax \def\eprint#1{#1}\fi

\bibitem{PSSb132-295}
\bibinfo{author}{\bibfnamefont{M.}~\bibnamefont{Inoue}},
  \bibinfo{author}{\bibfnamefont{M.}~\bibnamefont{Koyano}},
  \bibinfo{author}{\bibfnamefont{H.}~\bibnamefont{Negishi}},
  \bibinfo{author}{\bibfnamefont{Y.}~\bibnamefont{Ueda}}, \bibnamefont{and}
  \bibinfo{author}{\bibfnamefont{H.}~\bibnamefont{Sato}},
  \bibinfo{journal}{Phys. stat. solidi (b)}
  \textbf{\bibinfo{volume}{132}}(\bibinfo{number}{1}), \bibinfo{pages}{295}
  (\bibinfo{year}{1985}).

\bibitem{PSSb138-357}
\bibinfo{author}{\bibfnamefont{M.}~\bibnamefont{Koyano}},
  \bibinfo{author}{\bibfnamefont{H.}~\bibnamefont{Negishi}},
  \bibinfo{author}{\bibfnamefont{Y.}~\bibnamefont{Ueda}},
  \bibinfo{author}{\bibfnamefont{M.}~\bibnamefont{Sasaki}}, \bibnamefont{and}
  \bibinfo{author}{\bibfnamefont{M.}~\bibnamefont{Inoue}},
  \bibinfo{journal}{Phys. stat. solidi (b)}
  \textbf{\bibinfo{volume}{138}}(\bibinfo{number}{2}), \bibinfo{pages}{357}
  (\bibinfo{year}{1986}).

\bibitem{PRB38-3676}
\bibinfo{author}{\bibfnamefont{A.}~\bibnamefont{Fujimori}},
  \bibinfo{author}{\bibfnamefont{S.}~\bibnamefont{Suga}},
  \bibinfo{author}{\bibfnamefont{H.}~\bibnamefont{Negishi}}, \bibnamefont{and}
  \bibinfo{author}{\bibfnamefont{M.}~\bibnamefont{Inoue}},
  \bibinfo{journal}{Phys.~Rev.~B}
  \textbf{\bibinfo{volume}{38}}(\bibinfo{number}{6}), \bibinfo{pages}{3676}
  (\bibinfo{year}{1988}).

\bibitem{AdvPh36-1}
\bibinfo{author}{\bibfnamefont{R.~H.} \bibnamefont{Friend}} \bibnamefont{and}
  \bibinfo{author}{\bibfnamefont{A.~D.} \bibnamefont{Yoffe}},
  \bibinfo{journal}{Adv. Phys.}
  \textbf{\bibinfo{volume}{36}}(\bibinfo{number}{1}), \bibinfo{pages}{1}
  (\bibinfo{year}{1987}).

\bibitem{Hibma82}
\bibinfo{author}{\bibfnamefont{T.}~\bibnamefont{Hibma}},
  \emph{\bibinfo{title}{Structural Aspects of Monovalent Cation Intercalates of
  Layered Dichalcogenides}} (\bibinfo{publisher}{Academic Press},
  \bibinfo{address}{London -- New York}, \bibinfo{year}{1982}), pp.
  \bibinfo{pages}{285--313}.

\bibitem{TaTa_JPSJ97}
\bibinfo{author}{\bibfnamefont{Y.}~\bibnamefont{Tazuke}} \bibnamefont{and}
  \bibinfo{author}{\bibfnamefont{T.}~\bibnamefont{Takeyama}},
  \bibinfo{journal}{J.~Phys.~Soc.~Japan}
  \textbf{\bibinfo{volume}{66}}(\bibinfo{number}{3}), \bibinfo{pages}{827}
  (\bibinfo{year}{1997}).

\bibitem{YSM_JPC87}
\bibinfo{author}{\bibfnamefont{T.}~\bibnamefont{Yamasaki}},
  \bibinfo{author}{\bibfnamefont{N.}~\bibnamefont{Suzuki}}, \bibnamefont{and}
  \bibinfo{author}{\bibfnamefont{K.}~\bibnamefont{Motizuki}},
  \bibinfo{journal}{J.~Phys.~C: Solid State Phys.}
  \textbf{\bibinfo{volume}{20}}(\bibinfo{number}{3}), \bibinfo{pages}{395}
  (\bibinfo{year}{1987}).

\bibitem{JSSC17-9}
\bibinfo{author}{\bibfnamefont{Y.}~\bibnamefont{Arnaud}},
  \bibinfo{author}{\bibfnamefont{M.}~\bibnamefont{Chevreton}},
  \bibinfo{author}{\bibfnamefont{A.}~\bibnamefont{Ahouandjinou}},
  \bibinfo{author}{\bibfnamefont{M.}~\bibnamefont{Danot}}, \bibnamefont{and}
  \bibinfo{author}{\bibfnamefont{J.}~\bibnamefont{Rouxel}},
  \bibinfo{journal}{J. Solid State Chem.}
  \textbf{\bibinfo{volume}{18}}(\bibinfo{number}{1}), \bibinfo{pages}{9}
  (\bibinfo{year}{1976}).

\bibitem{FTT39-1618}
\bibinfo{author}{\bibfnamefont{V.~G.} \bibnamefont{Pleshchev}},
  \bibinfo{author}{\bibfnamefont{A.~N.} \bibnamefont{Titov}}, \bibnamefont{and}
  \bibinfo{author}{\bibfnamefont{A.~V.} \bibnamefont{Kuranov}},
  \bibinfo{journal}{Phys. Solid State}
  \textbf{\bibinfo{volume}{39}}(\bibinfo{number}{9}), \bibinfo{pages}{1442}
  (\bibinfo{year}{1997}). 

\bibitem{FMM81-629}
\bibinfo{author}{\bibfnamefont{A.~N.} \bibnamefont{Titov}},
  \bibinfo{journal}{Phys. Metals and Metallography}
  \textbf{\bibinfo{volume}{81}}(\bibinfo{number}{6}), \bibinfo{pages}{629}
  (\bibinfo{year}{1996}). 

\bibitem{Titov_MCLC98}
\bibinfo{author}{\bibfnamefont{A.}~\bibnamefont{Titov}},
  \bibinfo{author}{\bibfnamefont{S.}~\bibnamefont{Titova}},
  \bibinfo{author}{\bibfnamefont{M.}~\bibnamefont{Neumann}},
  \bibinfo{author}{\bibfnamefont{V.}~\bibnamefont{Pleschov}},
  \bibinfo{author}{\bibfnamefont{Y.}~\bibnamefont{Yarmoshenko}},
  \bibinfo{author}{\bibfnamefont{L.}~\bibnamefont{Krasavin}},
  \bibinfo{author}{\bibfnamefont{A.}~\bibnamefont{Dolgoshein}},
  \bibnamefont{and} \bibinfo{author}{\bibfnamefont{A.}~\bibnamefont{Kuranov}},
  \bibinfo{journal}{Mol. Cryst. Liq. Cryst.} \textbf{\bibinfo{volume}{311}},
  \bibinfo{pages}{161} (\bibinfo{year}{1998}).

\bibitem{PSSb86-11}
\bibinfo{author}{\bibfnamefont{J.~A.} \bibnamefont{Wilson}},
  \bibinfo{journal}{Phys stat. solidi (b)}
  \textbf{\bibinfo{volume}{1}}(\bibinfo{number}{1}), \bibinfo{pages}{11}
  (\bibinfo{year}{1978}).

\bibitem{JMP5-1117}
\bibinfo{author}{\bibfnamefont{M.~F.} \bibnamefont{Sykes}} \bibnamefont{and}
  \bibinfo{author}{\bibfnamefont{J.~W.} \bibnamefont{Essam}},
  \bibinfo{journal}{J. Math. Phys.}
  \textbf{\bibinfo{volume}{5}}(\bibinfo{number}{8}), \bibinfo{pages}{1117}
  (\bibinfo{year}{1964}).

\bibitem{Liang_NATO84}
\bibinfo{author}{\bibfnamefont{W.~Y.} \bibnamefont{Liang}}, in
  \emph{\bibinfo{booktitle}{Physics and Chemistry of Electrons and Ions in
  Condensed Matter. Proceedings of the NATO Advanced Study Institute}}, edited
  by \bibinfo{editor}{\bibfnamefont{J.~V.} \bibnamefont{Acrivos}},
  \bibinfo{editor}{\bibfnamefont{N.~F.} \bibnamefont{Mott}}, \bibnamefont{and}
  \bibinfo{editor}{\bibfnamefont{A.~D.} \bibnamefont{Yoffe}}
  (\bibinfo{publisher}{Reidel}, \bibinfo{address}{Dordrecht, Netherlands},
  \bibinfo{year}{1984}), pp. \bibinfo{pages}{459--478}. 

\bibitem{XPS_Handbook}
\bibinfo{author}{\bibfnamefont{J.~F.} \bibnamefont{Moulder}},
  \bibinfo{author}{\bibfnamefont{W.~F.} \bibnamefont{Stickle}},
  \bibinfo{author}{\bibfnamefont{P.~E.} \bibnamefont{Sobol}}, \bibnamefont{and}
  \bibinfo{author}{\bibfnamefont{K.~D.} \bibnamefont{Bomben}},
  \emph{\bibinfo{title}{Handbook of {X}-ray Photoelectron Spectroscopy}}
  (\bibinfo{publisher}{Perkin-Elmer Corporation}, \bibinfo{address}{Eden
  Prairie, Minnesota}, \bibinfo{year}{1992}). 

\bibitem{comment_MTiS2}
\bibinfo{note}{It is noteworthy that the corresponding spectra of
  $3d$-intercalated titanium disulphides exhibit an opposite trend: the Ti$2p$
  binding energy decreases on intercalation, consistently with other physical
  properties of $M_x$TiS$_2$ that may be well cast into a rigid-band model, see
  Ref.~\cite{PSSb132-295,PSSb138-357}. In TiS$_2$, the lattice is more rigid
  along the hexagonal axis direction (consistently with smaller width of the
  van der Waals gap) as compared to TiSe$_2$. Consequently, the formation of
  covalent bonds between titanium and intercalant atoms is more difficult in
  $M_x$TiS$_2$; the electrons provided by the dopant enter the ``rigid''
  conduction band. 
  %The peculiarity of the $M$TiSe$_2$ systems is that 
  %a softer lattice favors the increase of covalency 
  %in the chemical bonding (see Refs.~\onlinecite{FTT39-1618,Titov_MCLC98}).
  }

\bibitem{Heus_EPJ98}
\bibinfo{author}{\bibfnamefont{Y.~M.} \bibnamefont{Yarmoshenko}},
  \bibinfo{author}{\bibfnamefont{M.~I.} \bibnamefont{Katsnelson}},
  \bibinfo{author}{\bibfnamefont{E.~I.} \bibnamefont{Schreder}},
  \bibinfo{author}{\bibfnamefont{E.~Z.} \bibnamefont{Kurmaev}},
  \bibinfo{author}{\bibfnamefont{A.}~\bibnamefont{{\'S}lebarski}},
  \bibinfo{author}{\bibfnamefont{S.}~\bibnamefont{Plogman}},
  \bibinfo{author}{\bibfnamefont{T.}~\bibnamefont{Schlath{\"o}lter}},
  \bibinfo{author}{\bibfnamefont{J.}~\bibnamefont{Braun}}, \bibnamefont{and}
  \bibinfo{author}{\bibfnamefont{M.}~\bibnamefont{Neumann}},
  \bibinfo{journal}{Eur.~Phys.~J.~B}
  \textbf{\bibinfo{volume}{2}}(\bibinfo{number}{1}), \bibinfo{pages}{1}
  (\bibinfo{year}{1998}).

\bibitem{PRB60-6428}
\bibinfo{author}{\bibfnamefont{S.}~\bibnamefont{Plogmann}},
  \bibinfo{author}{\bibfnamefont{T.}~\bibnamefont{Schlath{\"o}lter}},
  \bibinfo{author}{\bibfnamefont{J.}~\bibnamefont{Braun}},
  \bibinfo{author}{\bibfnamefont{M.}~\bibnamefont{Neumann}},
  \bibinfo{author}{\bibfnamefont{Y.~M.} \bibnamefont{Yarmoshenko}},
  \bibinfo{author}{\bibfnamefont{M.~V.} \bibnamefont{Yablonskikh}},
  \bibinfo{author}{\bibfnamefont{E.~I.} \bibnamefont{Shreder}},
  \bibinfo{author}{\bibfnamefont{E.~Z.} \bibnamefont{Kurmaev}},
  \bibinfo{author}{\bibfnamefont{A.}~\bibnamefont{Wrona}}, \bibnamefont{and}
  \bibinfo{author}{\bibfnamefont{A.}~\bibnamefont{{\'S}lebarski}},
  \bibinfo{journal}{Phys.~Rev.~B}
  \textbf{\bibinfo{volume}{60}}(\bibinfo{number}{9}), \bibinfo{pages}{6428}
  (\bibinfo{year}{1999}).

\bibitem{YeLin_ADNDT85}
\bibinfo{author}{\bibfnamefont{J.~J.} \bibnamefont{Yeh}} \bibnamefont{and}
  \bibinfo{author}{\bibfnamefont{I.}~\bibnamefont{Lindau}},
  \bibinfo{journal}{At. Data Nucl. Data Tables}
  \textbf{\bibinfo{volume}{32}}(\bibinfo{number}{1}), \bibinfo{pages}{1}
  (\bibinfo{year}{1985}).

\bibitem{FTT38-1709}
\bibinfo{author}{\bibfnamefont{A.~N.} \bibnamefont{Titov}},
  \bibinfo{journal}{Phys. Solid State}
  \textbf{\bibinfo{volume}{38}}(\bibinfo{number}{10}), \bibinfo{pages}{1709}
  (\bibinfo{year}{1996}). 

\bibitem{PRB54-4405}
\bibinfo{author}{\bibfnamefont{S.~M.} \bibnamefont{Butorin}},
  \bibinfo{author}{\bibfnamefont{J.-H.} \bibnamefont{Guo}},
  \bibinfo{author}{\bibfnamefont{M.}~\bibnamefont{Magnuson}},
  \bibinfo{author}{\bibfnamefont{P.}~\bibnamefont{Kuiper}}, \bibnamefont{and}
  \bibinfo{author}{\bibfnamefont{J.}~\bibnamefont{Nordgren}},
  \bibinfo{journal}{Phys.~Rev.~B}
  \textbf{\bibinfo{volume}{54}}(\bibinfo{number}{7}), \bibinfo{pages}{4405}
  (\bibinfo{year}{1996}).

\bibitem{PRB55-4242}
\bibinfo{author}{\bibfnamefont{S.~M.} \bibnamefont{Butorin}},
  \bibinfo{author}{\bibfnamefont{J.-H.} \bibnamefont{Guo}},
  \bibinfo{author}{\bibfnamefont{M.}~\bibnamefont{Magnuson}}, \bibnamefont{and}
  \bibinfo{author}{\bibfnamefont{J.}~\bibnamefont{Nordgren}},
  \bibinfo{journal}{Phys.~Rev.~B}
  \textbf{\bibinfo{volume}{55}}(\bibinfo{number}{7}), \bibinfo{pages}{4242}
  (\bibinfo{year}{1997}).

\bibitem{Kawai_Cu_92}
\bibinfo{author}{\bibfnamefont{J.}~\bibnamefont{Kawai}},
  \bibinfo{author}{\bibfnamefont{K.}~\bibnamefont{Nakajima}},
  \bibinfo{author}{\bibfnamefont{K.}~\bibnamefont{Maeda}}, \bibnamefont{and}
  \bibinfo{author}{\bibfnamefont{Y.}~\bibnamefont{Goshi}},
  \bibinfo{journal}{Adv. X-ray Anal.} \textbf{\bibinfo{volume}{35}},
  \bibinfo{pages}{1107} (\bibinfo{year}{1992}).

\bibitem{PRB56-12238}
\bibinfo{author}{\bibfnamefont{M.}~\bibnamefont{Magnuson}},
  \bibinfo{author}{\bibfnamefont{N.}~\bibnamefont{Wassdahl}}, \bibnamefont{and}
  \bibinfo{author}{\bibfnamefont{J.}~\bibnamefont{Nordgren}},
  \bibinfo{journal}{Phys.~Rev.~B}
  \textbf{\bibinfo{volume}{56}}(\bibinfo{number}{19}), \bibinfo{pages}{12238}
  (\bibinfo{year}{1997}).

\bibitem{Heus_dichro}
\bibinfo{author}{\bibfnamefont{M.~V.} \bibnamefont{Yablonskikh}},
  \bibinfo{author}{\bibfnamefont{V.~I.} \bibnamefont{Grebennikov}},
  \bibinfo{author}{\bibfnamefont{Y.~M.} \bibnamefont{Yarmoshenko}},
  \bibinfo{author}{\bibfnamefont{E.~Z.} \bibnamefont{Kurmaev}},
  \bibinfo{author}{\bibfnamefont{S.~M.} \bibnamefont{Butorin}},
  \bibinfo{author}{\bibfnamefont{L.-C.} \bibnamefont{Duda}},
  \bibinfo{author}{\bibfnamefont{C.}~\bibnamefont{Sothe}},
  \bibinfo{author}{\bibfnamefont{T.}~\bibnamefont{K{\"a}{\"a}mbre}},
  \bibinfo{author}{\bibfnamefont{M.}~\bibnamefont{Magnuson}},
  \bibinfo{author}{\bibfnamefont{J.}~\bibnamefont{Nordgren}},
  \bibinfo{author}{\bibfnamefont{S.}~\bibnamefont{Plogmann}}, \bibnamefont{and}
  \bibinfo{author}{\bibfnamefont{M.}~\bibnamefont{Neumann}},
  \emph{\bibinfo{title}{Magnetic circular dichroism in {X}-ray fluorescence of
  {H}eusler alloys at threshold excitation}} \bibinfo{note}{(unpublished)}.

\bibitem{PRB94-5799}
\bibinfo{author}{\bibfnamefont{Y.}~\bibnamefont{Ma}},
  \bibinfo{journal}{Phys.~Rev.~B}
  \textbf{\bibinfo{volume}{49}}(\bibinfo{number}{9}), \bibinfo{pages}{5799}
  (\bibinfo{year}{1994}).

\bibitem{APA65-91}
\bibinfo{author}{\bibfnamefont{J.-E.} \bibnamefont{Rubensson}},
  \bibinfo{author}{\bibfnamefont{J.}~\bibnamefont{L{\"u}ning}},
  \bibinfo{author}{\bibfnamefont{S.}~\bibnamefont{Eisebitt}}, \bibnamefont{and}
  \bibinfo{author}{\bibfnamefont{W.}~\bibnamefont{Eberhardt}},
  \bibinfo{journal}{Appl. Phys. A} \textbf{\bibinfo{volume}{65}},
  \bibinfo{pages}{91} (\bibinfo{year}{1997}).

\bibitem{APA65-159}
\bibinfo{author}{\bibfnamefont{N.}~\bibnamefont{M{\aa}rtensson}},
  \bibinfo{author}{\bibfnamefont{M.}~\bibnamefont{Weinelt}},
  \bibinfo{author}{\bibfnamefont{O.}~\bibnamefont{Karis}},
  \bibinfo{author}{\bibfnamefont{M.}~\bibnamefont{Magnuson}},
  \bibinfo{author}{\bibfnamefont{N.}~\bibnamefont{Wassdahl}},
  \bibinfo{author}{\bibfnamefont{A.}~\bibnamefont{Nilsson}},
  \bibinfo{author}{\bibfnamefont{J.}~\bibnamefont{St{\"o}hr}},
  \bibnamefont{and} \bibinfo{author}{\bibfnamefont{M.}~\bibnamefont{Samant}},
  \bibinfo{journal}{Appl. Phys. A} \textbf{\bibinfo{volume}{65}},
  \bibinfo{pages}{159} (\bibinfo{year}{1997}).

\bibitem{EMRS99}
\bibinfo{author}{\bibfnamefont{A.~V.} \bibnamefont{Postnikov}},
  \bibinfo{author}{\bibfnamefont{M.}~\bibnamefont{Neumann}},
  \bibinfo{author}{\bibfnamefont{S.}~\bibnamefont{Plogmann}},
  \bibinfo{author}{\bibfnamefont{Y.~M.} \bibnamefont{Yarmoshenko}},
  \bibinfo{author}{\bibfnamefont{A.~N.} \bibnamefont{Titov}}, \bibnamefont{and}
  \bibinfo{author}{\bibfnamefont{A.~V.} \bibnamefont{Kuranov}},
  \emph{\bibinfo{title}{Magnetic properties of {$3d$}-doped {TiSe$_2$} and
  {TiTe$_2$} intercalate compounds}}, \bibinfo{note}{(to be published in
  Comput. Mater. Sci.)}.

\bibitem{wien97}
\bibinfo{author}{\bibfnamefont{P.}~\bibnamefont{Blaha}},
  \bibinfo{author}{\bibfnamefont{K.}~\bibnamefont{Schwarz}}, \bibnamefont{and}
  \bibinfo{author}{\bibfnamefont{J.}~\bibnamefont{Luitz}},
  \emph{\bibinfo{title}{{WIEN97}, {V}ienna {U}niversity of {T}echnology}}
  (\bibinfo{year}{1997}), \bibinfo{note}{improved and updated Unix version of
  the original copyrighted WIEN-code, which was published by P.~Blaha,
  K.~Schwarz, P.~Sorantin, and S.~B.~Trickey, in Comput. Phys. Commun. 59, 339
  (1990)}.

\bibitem{GGA_PBE_96}
\bibinfo{author}{\bibfnamefont{J.~P.} \bibnamefont{Perdew}},
  \bibinfo{author}{\bibfnamefont{K.}~\bibnamefont{Burke}}, \bibnamefont{and}
  \bibinfo{author}{\bibfnamefont{M.}~\bibnamefont{Ernzerhof}},
  \bibinfo{journal}{Phys. Rev. Lett.}
  \textbf{\bibinfo{volume}{77}}(\bibinfo{number}{18}), \bibinfo{pages}{3865}
  (\bibinfo{year}{1996}); 
\bibinfo{author}{\bibfnamefont{J.~P.} \bibnamefont{Perdew}},
  \bibinfo{author}{\bibfnamefont{K.}~\bibnamefont{Burke}}, \bibnamefont{and}
  \bibinfo{author}{\bibfnamefont{M.}~\bibnamefont{Ernzerhof}},
  \bibinfo{journal}{Phys. Rev. Lett.}
  \textbf{\bibinfo{volume}{78}}(\bibinfo{number}{7}), \bibinfo{pages}{1396}
  (\bibinfo{year}{1997});
\bibinfo{author}{\bibfnamefont{Y.}~\bibnamefont{Zhang}} \bibnamefont{and}
  \bibinfo{author}{\bibfnamefont{W.}~\bibnamefont{Yang}},
  \bibinfo{journal}{Phys. Rev. Lett.}
  \textbf{\bibinfo{volume}{80}}(\bibinfo{number}{4}), \bibinfo{pages}{890}
  (\bibinfo{year}{1998});
\bibinfo{author}{\bibfnamefont{J.~P.} \bibnamefont{Perdew}},
  \bibinfo{author}{\bibfnamefont{K.}~\bibnamefont{Burke}}, \bibnamefont{and}
  \bibinfo{author}{\bibfnamefont{M.}~\bibnamefont{Ernzerhof}},
  \bibinfo{journal}{Phys. Rev. Lett.}
  \textbf{\bibinfo{volume}{80}}(\bibinfo{number}{4}), \bibinfo{pages}{891}
  (\bibinfo{year}{1998}).

\bibitem{PRB51-17965}
\bibinfo{author}{\bibfnamefont{A.}~\bibnamefont{Leventi-Peetz}},
  \bibinfo{author}{\bibfnamefont{E.~E.} \bibnamefont{Krasovskii}},
  \bibnamefont{and} \bibinfo{author}{\bibfnamefont{W.}~\bibnamefont{Schattke}},
  \bibinfo{journal}{Phys.~Rev.~B}
  \textbf{\bibinfo{volume}{51}}(\bibinfo{number}{24}), \bibinfo{pages}{17965}
  (\bibinfo{year}{1995}).

\bibitem{TSRNC_98}
\bibinfo{author}{\bibfnamefont{S.}~\bibnamefont{Tinte}},
  \bibinfo{author}{\bibfnamefont{M.~G.} \bibnamefont{Stachiotti}},
  \bibinfo{author}{\bibfnamefont{C.~O.} \bibnamefont{Rodriguez}},
  \bibinfo{author}{\bibfnamefont{D.~L.} \bibnamefont{Novikov}},
  \bibnamefont{and} \bibinfo{author}{\bibfnamefont{N.~E.}
  \bibnamefont{Christensen}}, \bibinfo{journal}{Phys. Rev. B}
  \textbf{\bibinfo{volume}{58}}(\bibinfo{number}{18}), \bibinfo{pages}{11959}
  (\bibinfo{year}{1998}).

\bibitem{SYM_JMMM87}
\bibinfo{author}{\bibfnamefont{N.}~\bibnamefont{Suzuki}},
  \bibinfo{author}{\bibfnamefont{T.}~\bibnamefont{Yamasaki}}, \bibnamefont{and}
  \bibinfo{author}{\bibfnamefont{K.}~\bibnamefont{Motizuki}},
  \bibinfo{journal}{J.~Magn.~Magn.~Mater.}
  \textbf{\bibinfo{volume}{70}}(\bibinfo{number}{1--3}), \bibinfo{pages}{64}
  (\bibinfo{year}{1987}).

\bibitem{SYM_JPSJ89}
\bibinfo{author}{\bibfnamefont{N.}~\bibnamefont{Suzuki}},
  \bibinfo{author}{\bibfnamefont{T.}~\bibnamefont{Yamasaki}}, \bibnamefont{and}
  \bibinfo{author}{\bibfnamefont{K.}~\bibnamefont{Motizuki}},
  \bibinfo{journal}{J.~Phys.~Soc.~Japan}
  \textbf{\bibinfo{volume}{58}}(\bibinfo{number}{9}), \bibinfo{pages}{3280}
  (\bibinfo{year}{1989}).

\bibitem{YSKim_JJAP98}
\bibinfo{author}{\bibfnamefont{Y.-S.} \bibnamefont{Kim}},
  \bibinfo{author}{\bibfnamefont{M.}~\bibnamefont{Mizuno}},
  \bibinfo{author}{\bibfnamefont{I.}~\bibnamefont{Tanaka}}, \bibnamefont{and}
  \bibinfo{author}{\bibfnamefont{H.}~\bibnamefont{Adachi}},
  \bibinfo{journal}{Jpn. J. Appl. Phys.} \textbf{\bibinfo{volume}{37, Pt.
  1}}(\bibinfo{number}{9A}), \bibinfo{pages}{4878} (\bibinfo{year}{1998}).

\bibitem{Ovchin}
\bibinfo{author}{\bibfnamefont{A.~N.} \bibnamefont{Titov}},
  \bibinfo{author}{\bibfnamefont{A.~S.} \bibnamefont{Ovchinnikov}},
  \bibnamefont{and} \bibinfo{author}{\bibfnamefont{A.~V.}
  \bibnamefont{Kuranov}}, \emph{\bibinfo{title}{The role of {Me$3d$-Ti$3d$}
  hybridization in the reduction of magnetic moments of intercalant ions
  {$Co^{3+}$} and {$Cr^{3+}$} in {TiSe$_2$}}} \bibinfo{note}{(unpublished)}.

\end{thebibliography}

%
% --- Tables ---
%
\begin{table}
\caption{Effective magnetic moment $\mu_{\mbox{\tiny eff.}}$
(in $\mu_{\mbox{\tiny B}}$) as measured 
for different intercalant concentrations}
\begin{tabular}{cdddd}
$x$=           & 0.10 & 0.20 & 0.25 & 0.33 \\
\hline
Co$_x$TiSe$_2$ & 2.67 & 2.38 & 2.24 & 1.64 \\
Cr$_x$TiSe$_2$ & 3.85 & 3.03 & 3.29 & 2.56 \\
\end{tabular}
\label{tab:mom}
\end{table}

\begin{table}
\caption{Internal coordinates in relaxed $M_{1/3}$TiSe$_2$ supercells}
\begin{tabular}{cr@{.}lr@{.}lr@{.}lr@{.}l}
$M$ & \multicolumn{2}{c}{$\Delta_1{\times}10^3$} & 
      \multicolumn{2}{c}{$\Delta_2{\times}10^3$} & 
      \multicolumn{2}{c}{$\Delta_3{\times}10^3$} & 
      \multicolumn{2}{c}{$\Delta_4{\times}10^3$} \\
\hline
%Cr &    0&26 & $-$1&89 & 0&47 & $-$1&13 \\
Cr &    0&26 &    1&89 & 0&47 & $-$1&13 \\
Co & $-$0&02 &    1&53 & 0&50 & $-$1&40 \\
\end{tabular}
\label{tab:relax}
\end{table}

%
% --- Figures -------------------------------------
%

\vspace*{1.0cm}

%------------------------------------------------------
\end{document}